# $MoS_2$ Nanoparticles Grown on Graphene: An Advanced Catalyst for Hydrogen Evolution Reaction

Yanguang Li, Hailiang Wang, Liming Xie, Yongye Liang, Guosong Hong, and Hongjie Dai*

*Department of Chemistry, Stanford University, Stanford, CA 94305, USA*

E-mail: hdai@stanford.edu

ABSTRACT: Advanced materials for electrocatalytic and photoelectrochemical water splitting are central to the area of renewable energy. Here, we developed a solvothermal synthesis of $MoS_2$ nanoparticles selectively on reduced graphene oxide (RGO) sheets suspended in solution. The resulting $MoS_2$/RGO hybrid material possessed nanoscopic few-layer $MoS_2$ structures with abundant exposed edges stacked onto graphene, in strong contrast to large aggregated $MoS_2$ particles grown freely in solution without GO. The $MoS_2$/RGO hybrid exhibited superior electrocatalytic activity in the hydrogen evolution reaction (HER) to other $MoS_2$ catalysts. A Tafel slope of ~ 41 mV/decade was measured for $MoS_2$ catalysts in HER for the first time, far exceeding the activity of previous $MoS_2$ owing to the abundant catalytic edge sites of $MoS_2$ nanoparticles and excellent electrical coupling to the underlying graphene network. The ~ 41 mV/decade Tafel slope suggested the Volmer-Heyrovsky mechanism for $MoS_2$ catalyzed HER, with electrochemical desorption of hydrogen as the rate-limiting step.

Hydrogen is being vigorously pursued as a future energy carrier in transition from the current hydrocarbon economy[1]. In particular, sustainable hydrogen production from water splitting has attracted growing attention[1-3]. An advanced catalyst for electrochemical hydrogen evolution reaction (HER) should reduce the overpotential, and consequently increase the efficiency of this important electrochemical process[3]. The most effective HER electrocatalysts are Pt group metals. It remains challenging to develop highly active HER catalysts based on materials that are more abundant at lower costs[4].

$MoS_2$ is a material that has been commonly investigated as a catalyst for hydrodesulfurization[5]. Recent work showed that $MoS_2$ was a promising electrocatalyst for HER. Both computational and experimental results confirmed that the HER activity stemmed from the sulfur edges of $MoS_2$ plates while their basal plane were catalytically inert[6-8]. As a result, nanosized $MoS_2$ with exposed edges should be more active in HER electrocatalysis than materials in bulk forms. Previously, $MoS_2$ catalysts supported on Au[7], activated carbon[6], carbon paper[8] or graphite[9] were prepared by physical vapor deposition or annealing of molybdate in $H_2S$. Various overpotentials (from ~0.1 to ~0.4 V)[10] and Tafel slopes (55-60 mV/decade[7] or >120 mV/ decade[8]) were reported. The mechanism or reaction pathways of HER with $MoS_2$ catalysts also remained inconclusive.

In recent years, our group has been developing synthesis of nanostructured metal oxide or hydroxide materials on graphene sheets, either with graphene on solid substrates or graphene oxide (GO) sheets stably suspended in solutions[11-15]. These metal oxide- or hydroxide-graphene hybrids are novel owing to chemical and electrical coupling effects and utilization of the high surface area and electrical conductance of graphene, leading to advanced materials for nanoelectronics[11], energy storage devices (including pseudocapacitors[13] and lithium ion batteries[14]) and catalysis[15]. Here, we report the first synthesis of $MoS_2$ on reduced graphene oxide (RGO) sheets, and demonstrate the high HER electrocatalytic activity of the resulting $MoS_2$/RGO hybrid with low overpotential and small Tafel slopes.

$MoS_2$/RGO hybrid was synthesized by a one-step solvothermal reaction of $(NH_4)_2MoS_4$ and hydrazine in an *N, N*-dimethylformamide (DMF) solution of mildly oxidized graphene oxide (GO, Figure S1)[14] at 200°C (Figure 1A, nominal C/Mo atomic ratio ~10, see Supporting Information for synthetic details). During this process, the $(NH_4)_2MoS_4$ precursor was reduced to $MoS_2$ on GO with mildly oxidized GO transformed to RGO by hydrazine reduction[16]. Figure 2A-B showed the scanning electron microscopy (SEM) images of the resulting $MoS_2$/RGO hybrid, in which the RGO sheets were uniformly decorated with $MoS_2$ nanoparticles. Transmission electron microscopy (TEM) image (Figure 2C) showed that most of the $MoS_2$ nanoparticles were lying flat on graphene, with some possessing folded edges exhibiting parallel lines corresponding to the different layers of $MoS_2$ sheets (layer number ~ 3-10, Figure 2C insert). High resolution TEM revealed hexagonal atomic lattices in the $MoS_2$ basal planes and abundant open-edges of the nanoparticles (Figure 2D).

The $MoS_2$/RGO hybrid was characterized X-ray diffraction (XRD) and the broad diffraction peaks (Figure 2E) indicated nanosized $MoS_2$ crystal domains with hexagonal structure (PDF# 771716). Raman spectroscopy revealed the characteristic peaks[17] from $MoS_2$ at 373 and 400 $cm^{-1}$, and D, G and 2D bands of graphene in the hybrid (Figure 2F). Uniform distribution of $MoS_2$ on RGO was confirmed by micro-Raman imaging of the two components in the hybrid deposited on a substrate (Figure S2). X-ray photoelectron spectrum (XPS) confirmed the reduction of GO to RGO and Mo(VI) to Mo(IV)[18] (Figure S3). Residual oxygen content in the hybrid was measured to be < 4 at% (Figure S3).

Importantly, GO sheets provided a novel substrate for the nucleation and subsequent growth of $MoS_2$. The growth of $MoS_2$ was found selective on GO (by microscopy and Raman imaging) with little free particle growth in solution. Selective growth on GO was attributed to interactions between functional groups on GO sheets and Mo precursors in a suitable solvent environment[12,14-15]. In strong contrast, in the absence of GO, the exact same synthesis method produced $MoS_2$ coalesced into 3D-like particles of various sizes (Figure 1D). The drastically morphological difference highlighted the important role of GO as a novel support material for mediating the growth of nanomaterials. It is also important to note that replacing DMF with $H_2O$ as the solvent only afforded two separated phases of $MoS_2$ particles and RGO sheets (Figure S4).

We investigated the electrocatalytic HER activities of our $MoS_2$/RGO hybrid material deposited on a glassy carbon electrode in 0.5 M $H_2SO_4$ solutions using a typical three-electrode setup (see Supporting Information for experimental details). As a reference point, we also measured a commercial Pt catalyst (20 wt% of Pt on Vulcan carbon black) with high HER catalytic performances (with a near zero overpotential). Polarization curve ($i$-$V$ plot) recorded with our $MoS_2$/RGO hybrid on glassy carbon electrodes showed a small overpotential of ~0.1 V for HER (Figure 3A), beyond which the cathodic current rose rapidly under more negative potentials. In sharp contrast, free $MoS_2$ particles or RGO alone exhibited little HER activity (Figure 3A). $MoS_2$ particles physically mixed with carbon black at a similar C:Mo ratio also showed inferior performance to $MoS_2$/RGO (Figure S5). Linear portions of the Tafel plots (Figure 3B) were fit to the Tafel equation ($\eta = b \log j + a$, where $\eta$ is the overpotential, $j$ is the current density, $b$ is the Tafel slope), yielding Tafel slopes of ~30, ~41 and ~94 mV/decade for Pt, $MoS_2$/RGO hybrid and free $MoS_2$ particles respectively.

The $MoS_2$/RGO hybrid catalyst was further evaluated by depositing onto carbon fiber paper at a higher loading of 1 mg/cm$^2$ for reaching high electrocatalytic HER currents and comparisons with literature data of $MoS_2$ catalysts with similar loadings (Figure 3C). At the same potential, the $MoS_2$/RGO hybrid catalyst afforded significantly higher (iR-corrected) HER current densities than previous $MoS_2$ catalysts[6-9].

Three possible reaction steps have been suggested for HER in acidic media[19], including a primary discharge step (Volmer reaction):

$$H_3O^+ + e^- \rightarrow H_{ads} + H_2O \qquad b = \frac{2.3RT}{\alpha F} \approx 120 \text{ mV} \qquad (1)$$

where $R$ is ideal gas constant, $T$ is temperature, $\alpha \sim 0.5$ is the symmetry coefficient[19] and $F$ is Faraday constant. This is followed either by an electrochemical desorption step (Heyrovsky reaction):

$$H_{ads} + H_3O^+ + e^- \rightarrow H_2 + H_2O \quad b = \frac{2.3RT}{(1+\alpha)F} \approx 40 \text{ mV} \qquad (2)$$

or the recombination step (Tafel reaction):

$$H_{ads} + H_{ads} \rightarrow H_2 \qquad b = \frac{2.3RT}{2F} \approx 30 \text{ mV} \qquad (3)$$

Tafel slope is the inherent property of a catalyst, and determined by the rate limiting step of HER. Determination and interpretation of Tafel slope are important to elucidation of the elementary steps involved. With a very high $H_{ads}$ coverage ($\theta_H \sim 1$), HER on Pt surface is known to proceed through the Volmer-Tafel mechanism (reactions 1 and 3), and the recombination step is the rate limiting step at low over-potentials, as attested by the measured Tafel slope of 30 mV/decade[19]. Unfortunately, reaction mechanism on $MoS_2$ remained inconclusive since its first HER study more than forty years ago[20]. Even though previous density functional theory (DFT) calculations suggested a $H_{ads}$ coverage of 0.25-0.50[6], which could favor an electrochemical desorption mechanism, experimental mechanistic studies were inconclusive due to the discrepancy in a wide range of HER Tafel slopes reported[7-8]. The observed Tafel slope of ~41 mV/decade in the current work is the smallest ever measured with $MoS_2$-based catalyst, suggesting the electrochemical desorption as the rate limiting step[19] and the Volmer-Heyrovsky HER mechanism (reactions 1 and 2) for HER catalyzed by the $MoS_2$/RGO hybrid.

We attribute the high performance of our $MoS_2$/RGO hybrid catalyst in HER to strong chemical and electronic coupling between GO sheets and $MoS_2$. Chemical coupling/interactions afforded selective growth of highly dispersed $MoS_2$ nanoparticles on GO free of aggregation. The small size and high dispersion of $MoS_2$ on GO afford abundant, accessible edges to serve as active catalytic sites for HER. Electrical coupling to the underlying graphene sheets in an interconnected conducting network afforded rapid electron transport from the less-conducting $MoS_2$ nanoparticles to the electrodes. To glean this effect, we performed impedance measurements at an overpotential of $\eta = 0.12$ (Figure S6). The $MoS_2$/RGO hybrid exhibited much lower impedance (Faradaic impedance $Z_f$, or charge transfer impedance ~ 250 $\Omega$[21]) than free $MoS_2$ particles ($Z_f$ ~10 K$\Omega$). The significantly reduced Faradaic impedance afforded markedly faster HER kinetics with the $MoS_2$/RGO hybrid catalyst.

Another important criterion for a good electrocatalyst is strong durability. To assess this, we cycled our MoS$_2$/RGO hybrid catalyst continuously for 1,000 times. At the end of cycling, the catalyst afforded similar *i-V* curves as before with negligible loss of the cathodic current (Figure 3D).

In conclusion, we synthesized MoS$_2$ nanoparticles on RGO sheets via a facile solvothermal approach. With highly exposed edges and excellent electrical coupling to the underlying graphene sheets, the MoS$_2$/RGO hybrid catalyst exhibited excellent HER activity with a small overpotential of ~0.1V, large cathodic currents and a Tafel slope down to 41 mV/decade. This was the smallest Tafel slope reported for MoS$_2$ catalysts, suggesting electrochemical desorption as the rate limiting step in the catalyzed HER. Thus, the approach of materials synthesis on graphene led to an advanced MoS$_2$ electro-catalyst with highly competitive performances among various HER electrocatalytic materials.

**Acknowledgment.** This work was supported partially by ONR and NSF CHE-0639053.

**Supporting Information Available:** Experimental procedures and supportive data. This information is available free of charge via the Internet at http://pubs.acs.org/.

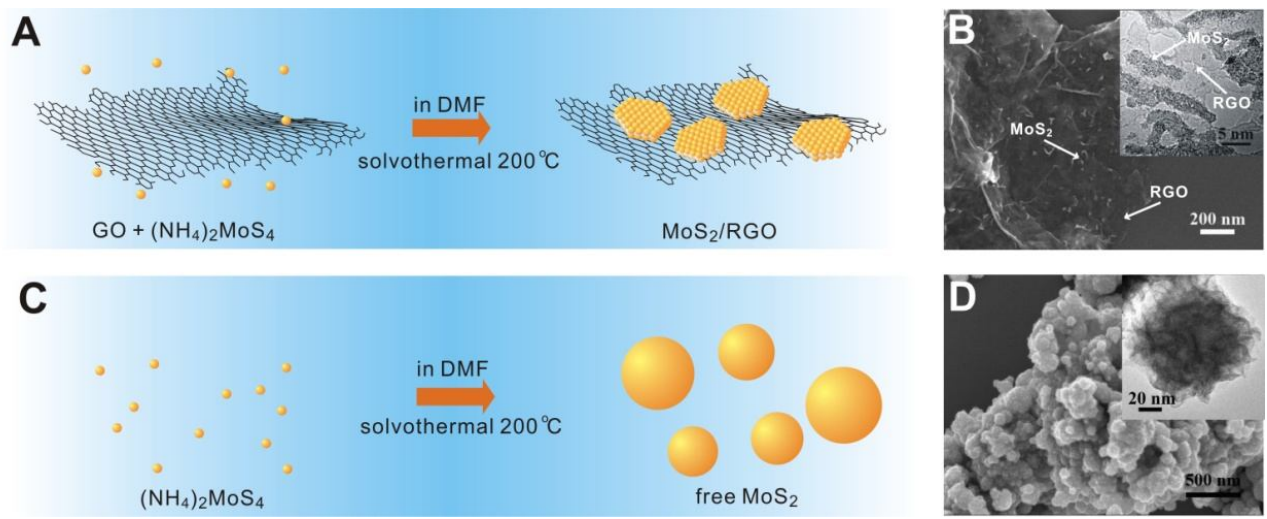

***Figure 1.*** Synthesis of $MoS_2$ in solution with and without graphene sheets. (A) Schematic solvothermal synthesis with GO sheets to afford $MoS_2$/RGO hybrid. (B) SEM and TEM (inserted) images of the $MoS_2$/RGO hybrid. (C) Schematic solvothermal synthesis without any GO sheets, which resulted in free large $MoS_2$ particles. (D) SEM and TEM (inserted) images of the free particles.

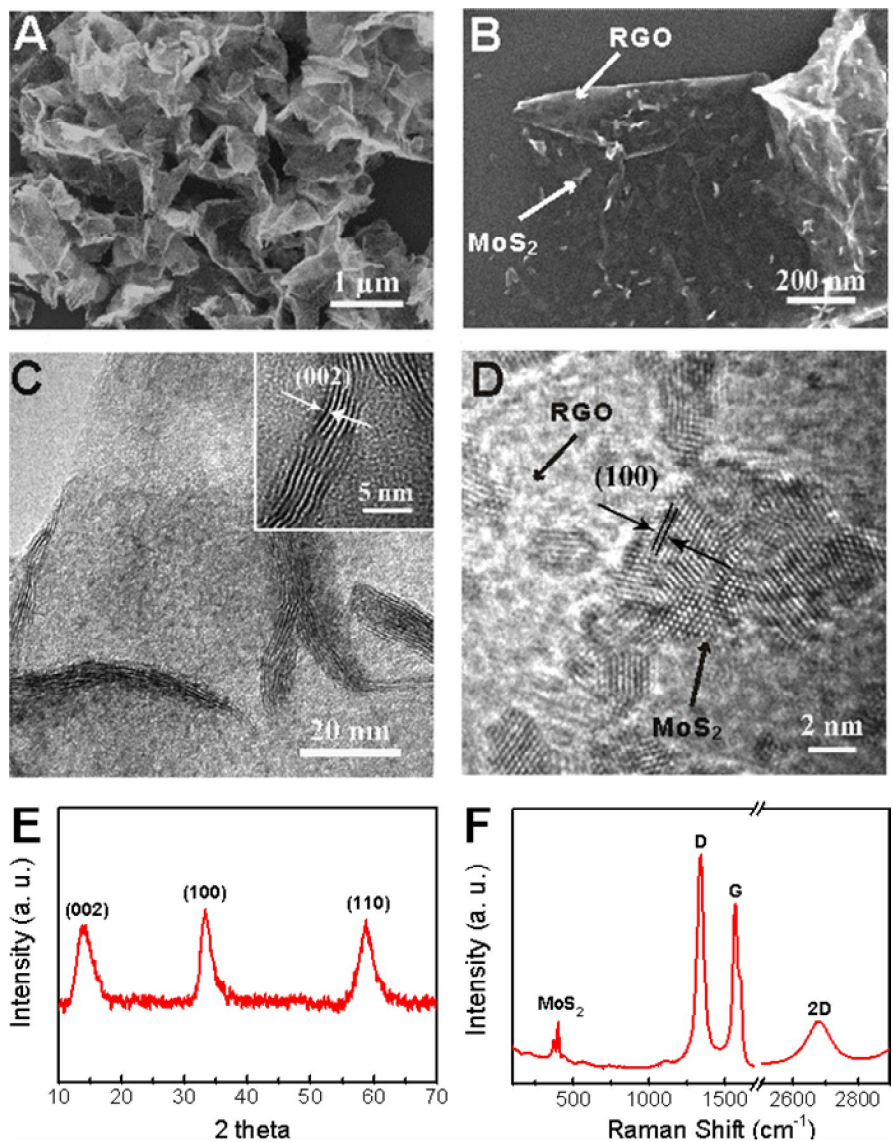

*Figure 2.* MoS$_2$ nanoparticles on graphene in the MoS$_2$/RGO hybrid. (A-B) SEM images of the MoS$_2$/RGO hybrid. (C) TEM image showing folded edges of MoS$_2$ particles on RGO in the hybrid. The insert magnifies the folded edge of a MoS$_2$ nanoparticle. (D) High resolution TEM image showing nanosized MoS$_2$ with highly exposed edges stacked on a RGO sheet. (E) XRD and (F) Raman spectrum of the hybrid.

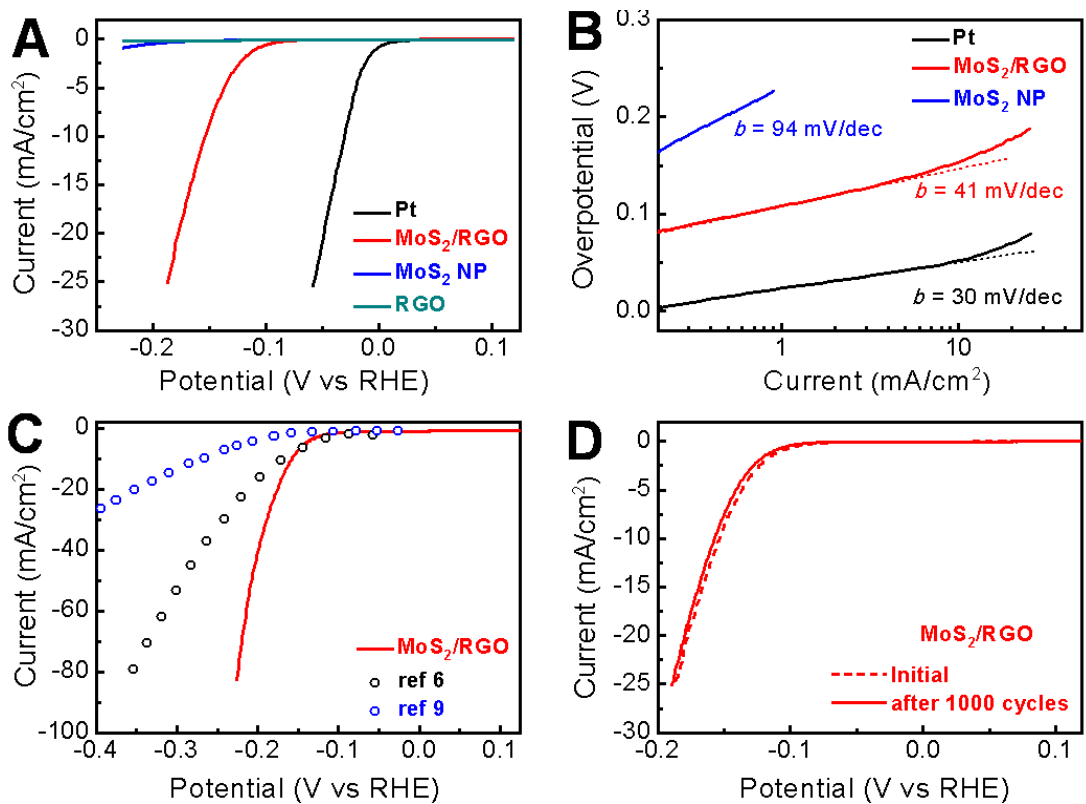

*Figure 3.* (A) Polarization curves obtained with several catalysts as indicated and (B) corresponding Tafel plots recorded on glassy carbon electrodes with a catalyst loading of 0.28 mg/cm$^2$. (C) Polarization curves recorded on carbon fiber paper with a loading of 1 mg/cm$^2$, in comparison with two literature results with similar catalyst loadings. (D) A durability test of the MoS$_2$/RGO hybrid catalyst. Negligible HER current was lost after 1000 cycles from -0.3 to +0.7 V at 100 mV/s.